\documentclass[nojss]{jss}\usepackage[]{graphicx}\usepackage[]{color}
\makeatletter
\def\maxwidth{ %
  \ifdim\Gin@nat@width>\linewidth
    \linewidth
  \else
    \Gin@nat@width
  \fi
}
\makeatother

\definecolor{fgcolor}{rgb}{0.345, 0.345, 0.345}

\usepackage{framed}
\makeatletter
 {\par\unskip\endMakeFramed%
 \at@end@of@kframe}
\makeatother

\definecolor{shadecolor}{rgb}{.97, .97, .97}
\definecolor{messagecolor}{rgb}{0, 0, 0}
\definecolor{warningcolor}{rgb}{1, 0, 1}
\definecolor{errorcolor}{rgb}{1, 0, 0}

\usepackage{alltt}

\usepackage{amsmath}
\usepackage{float}

\author{Nina Zumel\\Win-Vector LLC \And
        John Mount\\Win-Vector LLC}
\title{\pkg{vtreat}: a \code{data.frame} Processor for Predictive Modeling
}

\Plainauthor{Nina Zumel, John Mount} 
\Plaintitle{vtreat: a data.frame Processor for Predictive Modeling} 
\Shorttitle{\pkg{vtreat}: a data.frame Processor for Predictive Modeling} 

\Abstract{
In this paper, we will look at common problems that arise in data that is used for supervised machine learning or predictive modeling tasks, and describe how to address them with the \pkg{vtreat} \proglang{R} package.  \pkg{vtreat} prepares real-world data for predictive modeling in a reproducible and statistically sound manner, and is a valuable addition to the data science work-flow.
}
\Plainkeywords{vtreat, data preparation, nested models, predictive modeling, classification, regression, R} 

\Submitdate{2016-11-07}

\Address{
  Nina Zumel\\
  Win-Vector LLC\\
  552 Melrose Ave.\\
  San Francisco, Ca 94127\\
  E-mail: \email{nzumel@win-vector.com}\\
  URL: \url{http://win-vector.com/}

  John Mount\\
  Win-Vector LLC\\
  552 Melrose Ave.\\
  San Francisco, Ca 94127\\
  E-mail: \email{jmount@win-vector.com}\\
  URL: \url{http://win-vector.com/}
}

\IfFileExists{upquote.sty}{\usepackage{upquote}}{}
\begin{document}


\section{Introduction}
\label{sec:intro}


\subsection[The vtreat package]{The \pkg{vtreat} package}

\pkg{vtreat} is an 
\code{data.frame} processor or conditioner that prepares real-world data
for predictive modeling in a statistically sound manner.
\pkg{vtreat} is available for both \proglang{R} (\cite{Rref}) \cite{vtreatcran} and \proglang{Python} \cite{vtreatpython}\footnote{
  This article will describe only the \proglang{R} version of the package.  Documentation and examples
  for the Python version can be found here: \url{https://github.com/WinVector/pyvtreat}.}
We call this step or process
``data treatment'', or ``data conditioning.''

The package's function is to collect statistics on a
\code{data.frame} in order to produce a \textit{treatment plan}.
This treatment plan is then used to process subsequent \code{data.frame}s for model training
and model application. This processed data frame is meant to be as useful for
predictive modeling as the original and be easier to work with:
having no missing values, and no string/factor/categorical values.
\pkg{vtreat} serves as a powerful alternative to
\code{model.matrix}, which is implicitly used in many \proglang{R} modeling tasks.

The purpose of this article is to specify, document, and justify these procedures.

\subsection{The problem}

Even with modern machine learning techniques and standard statistical methods, there are common correctable data issues that can cause modeling to fail. Typical treatable data problems include:

\begin{itemize}
\item Missing or invalid values (either in numeric or categorical variables).
\item Novel levels discovered in categorical variables during model application.
\item High cardinality categorical variables, which can have both statistical issues, and operational issues\footnote{Such as the \pkg{randomForest} package's limit of 63 levels for categorical variables}.
\item Wide data: having too many candidate variables (often a symptom of under-curated data sets).
\end{itemize}

\pkg{vtreat} automates the mitigation of these issues, which we call data treatment.  The goal of \pkg{vtreat} is to reliably generate an \proglang{R} \code{data.frame} that is safe to work with, as:

\begin{itemize}
\item Missing or invalid values are replaced with safe valid values, and further indicated by additional dummy variables.
\item Categorical variables are represented in a manner that is robust to the appearance of novel levels during model application.
\item High cardinality categorical values are safely converted to numerical \textit{impact codes}\footnote{Also called \textit{effect codes}, and more recently \textit{target coding}.} while avoiding introducing nested model bias (defined in Section~\ref{sec:nestedmodelbiastheory}).
\item Non-rare levels of all categorical variables are also retained as explicit indicator variables (an advantage for many modeling techniques, such as trees and other recursive partitioning methods).
\item Estimated variable significances are supplied for user-controlled variable pruning.
\end{itemize}

\pkg{vtreat} is designed to prepare data for predictive modeling (the use of sets of variables to estimate an outcome).
This application choice allows the primary data treatment strategy to transform the original variables into multiple derived columns.\footnote{A strategy that may be less appropriate for statistical
inference problems which seek to relate effects to original variables or columns.}
The use of \pkg{vtreat} lowers the required amount of ad-hoc per-project data cleaning effort
and procedure documentation
by supplying a specific, \textit{citable} treatment implementation.

\subsection[vtreat Design Principles]{\pkg{vtreat} Design Principles}

The set of transformations we are documenting are those we have found useful in what we call a predictive modeling context.
This the assumption that the data is being prepared for a ``black box'' predictive modeling task (such as regression or classification)
in the machine learning sense.  We are not claiming these are the appropriate transformations for 
visualization, reporting, causal inference, or coefficient inference tasks.
Limitations of our approach are noted throughout this article, and
summarized in Section~\ref{sec:limitations}.

We outline some of our design principles below.

\subsubsection{``Not a domain expert'' assumption}

\pkg{vtreat} avoids any transformation that cannot be reliably performed without domain expertise.

For example \pkg{vtreat} does not perform outlier detection or density
estimation to attempt to discover \textit{sentinel values} hidden in
numeric data.  We consider reliably detecting such values (which can
in fact ruin an analysis when not detected) a domain specific question.  To
get special treatment of such values the analyst needs to first
convert them to separate indicators and/or a special value
such as \code{NA}.

This is also why \pkg{vtreat} does not default
to collaring or Winsorizing numeric values (restricting numeric values
to ranges observed during treatment design).  For some variables
Winsorizing seems harmless, for others (such as time) it is a catastrophe.
This determination
can be subjective, which is why we include the feature as a user control.

\subsubsection{``Not the last step'' assumption}

One of the design principles of \pkg{vtreat} is the assumption that any use of \pkg{vtreat} is followed
by a sophisticated modeling technique.  That is: a later technique that
can reason about groups of variables. So \pkg{vtreat} defers reasoning about groups of variables and other post-processing
to this technique.

This is one reason \pkg{vtreat} allows both level indicators and complex derived variables (such as effects or impact coded variables) to be taken from the same original categorical variable, even though this
can introduce linear dependency among derived variables.  \code{vtreat} does prohibit constant or non-varying derived variables
as those are traditionally considered anathema in modeling.

\proglang{R}'s base \code{lm} and \code{glm(family=binomial)} methods are sophisticated in that they do
work properly in the presence of co-linear independent variables, as both methods automatically
remove a set of redundant variables during analysis.  However, in general we would recommend regularized techniques
as found in \pkg{glmnet} as a defense against near-dependency among variables.

\pkg{vtreat} variables are intended to be used with regularized statistical methods, which is one reason that
for categorical variables no value is picked as a reference level to build contrasts.  For $L2$ or Tikhonov regularization
it can be more appropriate to regularize indicator-driven effects towards zero than towards a given reference level.

This is also one reason the user must supply a variable pruning significance; the variable pruning level is sensitive to the modeling goals,
number of variables, and number of training examples.  Variable pruning is so critical in industrial data science practice
we feel we must supply some tools for it, but also must leave the control to the user.  
Any joint dimension reduction technique (other than variable pruning)
is again left as a next step\footnote{Though \pkg{vtreat}'s scaling feature can be a useful preparation for principal components analysis, please see \cite{yawarescaling}.}

\pkg{vtreat}'s explicit indication of missing values is meant to allow the next stream processing to use missingness
as possibly being informative and work around the limitations of \pkg{vtreat}'s simple unconditioned point replacement of missing values.

\subsubsection{``Consistent estimators'' principle}

The estimates \pkg{vtreat} returns should be consistent in the sense that
they converge to ideal non-constant values as the amount of data
available for calibration or design goes to infinity.\footnote{Note this is a 
  colloquial sense of ``consistent'', not the formal statistical term denoting 
  convergence in probability.}
This means we
can have introduce derived variables that are expectations (such as \code{catB} and \code{catN}
variables), prevelances or frequencies (such as \code{catP}), and even
conditional deviations (such as \code{catD}).  The principle
forbids other tempting summaries such as conditional counts (which
scale with data) or conditional significances (which when they converge, 
converge to the non-informative constant zero).

\subsection{Related work}

Statistically valid data treatment in the service of
predictive modeling is an under-served topic.  This article is largely
a codification of the authors' original work on this topic.  The ideas have
precedent (especially y-aware re-encoding high cardinality categorical
variables) but are not often
fully described.

Our preparing for predictive modeling emphasis differs from related works
such as: data wrangling/shaping (as demonstrated in
\cite{RforDataScience}), model training control (as demonstrated in
\cite{caret}, \cite{JMLR:v17:15-066}, and \cite{slcran}), systematic missing
value imputation (\cite{Kabacoff} chapter 18), and specialized
variable transforms/scaling\footnote{\code{caret::preProcess} does
supply missing value imputation for \textit{numeric} predictors, but
its primary purpose is transformation, centering, and scaling
independent of a declared modeling target.}.

Related works directly addressing data cleaning include
\cite{appliedmr}, 
\cite{introToDataCleaning},  \cite{dataprep}, and our own \cite{PDSwR}.

Additional practitioner oriented references include: \cite{DataPrepDM}, \cite{EDMDC}, 
\cite{DataWarehouseETL}, \cite{SASclean}, 
\cite{bpDataCleaning}, \cite{BadData}, and \cite{CleanData}.

\subsection{Outline}

We organize the paper as follows: Section~\ref{sec:theory} covers the
problems with preparing data for predictive modeling, and the
principles behind \pkg{vtreat}'s design.
Section~\ref{sec:basicfunctions} documents the operational
aspects of the \pkg{vtreat} implementation. We discuss the limitations of the current \pkg{vtreat} implementation
in Section~\ref{sec:limitations} and conclude in Section~\ref{sec:finalRemarks}.

\section{Principles of data preparation for predictive modeling}
\label{sec:theory}

In this section we present a number of the common data problems that \pkg{vtreat} addresses. We discuss the principles behind the various solutions to these problems, and motivate our choices for \pkg{vtreat}'s chosen approach\footnote{Further discussion on the principles of data preparation for predictive modeling can be found in
\cite{PDSwR} and \cite{dataprep}.}.

\subsection{Handling missing and bad values in data}

Bad values can stop an analysis in its tracks.  Such
values can be missing (\code{NA}) or problematic types (\code{NaN, Inf}). They can also be invalid values: invalid category levels, implausible numeric values, or sentinel values (a value used to represent ``unknown'' or ``not applicable'' or other special cases in numeric data). When not addressed, bad values can lead to invalid or poorly predicting models through the inadvertent removal of training data, or misleading input to the modeling algorithm.

Identifying bad values often requires domain knowledge of the plausible values of the variables. For the purposes of this discussion, we assume that invalid categoric or numeric values (including sentinels) have been detected and converted to \code{NA}.

If the number of missing values is small, it may be safe to simply drop those rows during training. In many cases, however, there may be a substantial number of rows with missing values, so that dropping them can lead to invalid analyses. This is especially true when there are additional distributional differences between the dropped and retained data. Also, production models usually need to score all observations, even when those observations have missing values.

\code{NA}s in categorical variables can be treated as an additional category level. The appropriate treatment of \code{NA}s of course depends on why the data is missing. Generally, we consider values to be either \emph{missing at random} or \emph{missing systematically}.

\subsubsection{When values are missing randomly}

Consider the following dataset of sleep statistics for various animals:

\begin{Schunk}
\begin{Sinput}
R> library("ggplot2")
R> data("msleep")
R> str(msleep, width = 70, strict.width = "cut")
\end{Sinput}
\begin{Soutput}
Classes 'tbl_df', 'tbl' and 'data.frame':	83 obs. of  11 variables:
 $ name        : chr  "Cheetah" "Owl monkey" "Mountain beaver" "Gre"..
 $ genus       : chr  "Acinonyx" "Aotus" "Aplodontia" "Blarina" ...
 $ vore        : chr  "carni" "omni" "herbi" "omni" ...
 $ order       : chr  "Carnivora" "Primates" "Rodentia" "Soricomorp"..
 $ conservation: chr  "lc" NA "nt" "lc" ...
 $ sleep_total : num  12.1 17 14.4 14.9 4 14.4 8.7 7 10.1 3 ...
 $ sleep_rem   : num  NA 1.8 2.4 2.3 0.7 2.2 1.4 NA 2.9 NA ...
 $ sleep_cycle : num  NA NA NA 0.133 0.667 ...
 $ awake       : num  11.9 7 9.6 9.1 20 9.6 15.3 17 13.9 21 ...
 $ brainwt     : num  NA 0.0155 NA 0.00029 0.423 NA NA NA 0.07 0.098..
 $ bodywt      : num  50 0.48 1.35 0.019 600 ...
\end{Soutput}
\end{Schunk}

There are several missing measurements for \code{brainwt} and some of the sleep statistics. It is possible that this data is missing because of a faulty sensor -- in other words, the data collection failed at random (independently of the value censored, and all other variables or outcomes). In this case, one can replace the missing values with stand-ins, such as inferred values, distributions of values, or the expected or mean value of the nonmissing data. Assuming that the rows with missing values are distributed the same way as the others, this estimate will be correct on average, and is an easy fix to implement. This estimate can be improved when missing values are related to other variables in the  data: for instance, brain weight may be related to body weight. Note that the method of imputing a missing value of an input variable based on the other input variables can be applied to categorical data as well. \cite{Kabacoff} includes an extensive discussion of several methods for imputing missing values that are available in \proglang{R}.

\subsubsection{When values are missing systematically}

Replacing missing values by the mean, as well as many more sophisticated methods for imputing missing values, assumes that the rows with missing data are in some sense random (the faulty sensor situation). It's possible that the rows with missing data are systematically different from the others.  For example, it may be difficult to measure REM sleep for animals that are small, or don't sleep much; possibly some animals don't have REM sleep. In these situations (particularly the last) imputing missing values using one of the preceeding methods is not appropriate.  In this situation, a practical solution is to fill in the missing values with a nominal value, perhaps either the mean value of the nonmissing data or zero, and additionally to add a new variable that tracks which data have been altered.\footnote{This method is seen in \cite{appliedmr}.}  This could be achieved by code such as the following:

\begin{Schunk}
\begin{Sinput}
R> msleep$sleep_rem_isBAD <- is.na(msleep$sleep_rem)
R> msleep$sleep_rem <- ifelse(msleep$sleep_rem_isBAD, 
+                             mean(msleep$sleep_rem, na.rm=TRUE), 
+  		           msleep$sleep_rem)
\end{Sinput}
\end{Schunk}

For the motivation behind this approach, we look at linear models. Suppose we want to predict the outcome $y$ using (among other variables) the input $x$, which has missing values. If we fill in the missing $x$ values with \code{mean(x)}, we are essentially saying that $x$ has no net effect on $y$ when $x$ is missing. If we also add an additional indicator variable $x_{isBAD}$, then this indicator will allow a model to re-estimate the expected value of $y$ for those rows where $x$ is missing (conditioned on the values of any additional input variables). More complex machine learning algorithms might also be able to model nonlinear effects, such as interactions between missingness and other variables.

While this approach is not as statistically sophisticated as some of the imputation methods in \cite{Kabacoff}, it can be sufficient for a downstream machine learning algorithm to learn any relationship between rows with missing values and the outcome of interest.

When it is not known whether missing values in the data are missing randomly or systematically, it is safer and more conservative to assume, as \pkg{vtreat} does, that they may be missing systematically. For business data, this is the more likely scenario. Because missingness is often an indication of data provenance in a business setting, the missingness indicator column can be a highly informative variable -- sometimes more informative than the values of the original variable.

\subsection{Missing values in categorical variables}

For categorical variables one can treat \code{NA} or missing values as
just another standard level; \pkg{vtreat} takes this approach. If
\code{NA} occurs during variable treatment design, we get usable statistics on
the relationship between missingness and the outcome; if \code{NA} does not occur during treatment design
it is treated as a novel level, as discussed in Section~\ref{subsec:novel}.

\subsection{Novel categorical levels and indicators}\label{subsec:novel}
Unlike many programming languages commonly used for statistical modeling, most \proglang{R} modeling functions can accept categorical variables directly. While this has many advantages, one downside is that \proglang{R} models do not gracefully handle data containing categorical levels that were not present in the training data.

\begin{Schunk}
\begin{Sinput}
R> df <- data.frame(x=c('a', 'a', 'b', 'b', 'c', 'c'), 
+                   y=1:6, 
+                   stringsAsFactors=FALSE)
R> model <- lm(y~x, data=df)
R> newdata <- data.frame(x=c('a', 'b', 'c', 'd'), 
+                        stringsAsFactors=FALSE)
R> tryCatch(
+     predict(model, newdata=newdata), 
+     error = function(e) print(strwrap(e)))
\end{Sinput}
\begin{Soutput}
[1] "Error in model.frame.default(Terms, newdata, na.action ="    
[2] "na.action, xlev = object$xlevels): factor x has new levels d"
\end{Soutput}
\end{Schunk}

To avoid this, one would like to detect novel levels in new data and encode them in a way that the model can understand. This task is much easier when representing categorical variables as indicators.

In \pkg{vtreat}, the procedure is as follows:

\begin{samepage} 
\begin{Schunk}
\begin{Sinput}
R> library("vtreat")
R> treatplan <- designTreatmentsN(df, 'x', 'y')
R> varnames <- treatplan$scoreFrame$varName[treatplan$scoreFrame$cod=="lev"]
R> newdata_treat <- prepare(treatplan, newdata, 
+                           pruneSig=NULL, varRestriction=varnames)
\end{Sinput}
\end{Schunk}
\end{samepage}

The function \code{designTreatmentsN} creates new derived variables -- including indicators --  from the original data. Indicator variables have the designation (or code) \code{lev} in the resulting treatment plan. The function \code{prepare} applies a treatment plan to a new data frame that has the same input columns as the original one. We give a detailed discussion of the \pkg{vtreat} workflow in Section~\ref{sec:basicfunctions}.

This process converts new data from its original encoding:

\begin{samepage} 
\begin{Schunk}
\begin{Sinput}
R> print(newdata)
\end{Sinput}
\begin{Soutput}
  x
1 a
2 b
3 c
4 d
\end{Soutput}
\end{Schunk}
\end{samepage}

Producing derived variables encoded as indicators, even in the presence of
novel levels:

\begin{samepage} 
\begin{Schunk}
\begin{Sinput}
R> print(newdata_treat)
\end{Sinput}
\begin{Soutput}
  x_lev_x_a x_lev_x_b x_lev_x_c
1         1         0         0
2         0         1         0
3         0         0         1
4         0         0         0
\end{Soutput}
\end{Schunk}
\end{samepage}

The resulting data can be safely input to a model that was trained on the original data set where the novel level `d' was not present.

\subsubsection{Representing novel levels}
How are novel levels best represented? For the purposes of this discussion, assume that in our training data the categorical variable $x$ takes on the values $a, b, c, d, e$ with the observed frequencies $f_a, f_b, f_c, f_d, f_e$, respectively. Further assume that $f_d, f_e \ll f_a, f_b, f_c$ -- that is, $d$ and $e$ are rare levels. We will represent a value of $x$ as the tuple $(s_a, s_b, s_c, s_d, s_e)$. Usually the components $s_i$ take on the values 0 or 1: for example the value $c$ is represented by the tuple $(0, 0, 1, 0, 0)$, and so on.

After we have fit a model to the training data, we apply it to new data, in which we observe $x$ take on the previously unseen value $w$. How do we represent $w$?  There are at least three possible solutions:

\begin{description}
\item[1. Novel levels are represented as ``no level''.] In other words:
\[
w \rightarrow (0, 0, 0, 0, 0).
\]
This is the most straightforward representation. In effect we assume that when $x$ takes on a previously unseen value, it has no effect on the outcome.

\item[2. Novel levels are weighted proportional to known levels.] In other words:
\[
w \rightarrow (f_a, f_b, f_c, f_d, f_e).
\]
This is analogous to the ``faulty sensor'' assumption for missing data, or better, a ``transcription error'': we in effect assume that the novel level is really one of the known levels, proportional to the prevalence of each level in the training data. A linear model would tend to predict the weighted average of the outcomes that would be predicted for each of the known levels.

\item[3. Novel levels are treated as uncertainty among rare levels.] In other words:
\[
w \rightarrow (0, 0, 0, 0.5, 0.5).
\]
(Recall that we consider the values $d$ and $e$ as ``rare''.) A variation on this is to pool the rare levels into a single category level, \emph{rare}, before modeling, and then re-encoding novel levels as \emph{rare} during model deployment. The intuition behind pooling is that previously unobserved values are simply rare, and that rare levels might behave somewhat similarly with respect to the output.
\end{description}

\pkg{vtreat} either uses the first approach, or uses an additional pooled rare indicator (similar in the third approach), choosing which approach by statistical test.

Each of these representations is best for different situations, and may return wildly inaccurate predictions on datums that manifest a novel level. On the other hand, the model can successfully accept unexpected values without crashing, and since any individual novel level is rare, it will not affect overall model performance much. A possible exception is high-cardinality categorical variables, which take on one of a very large number of possible values. In such cases, ``rare'' levels may not be so rare, in aggregate (an alternative way of putting this is that with such variables, ``most levels are rare''). We will discuss high-cardinality categorical variables in the next section.

\subsection{High-cardinality categorical variables}

Another type of problematic variable is the categorical variable with many possible values (or levels, in \proglang{R} parlance): zip codes and business codes like NAICS codes fall into this category. Such high-cardinality variables can cause issues for two reasons.

First, computationally speaking, a categorical variable with $k$ levels is treated by most machine learning algorithms as the equivalent of $k-1$ numerical (0/1) variables. For example, suppose a U.S.-based market researcher wanted to build a nationwide customer model that used geographic information about the customer, including the zip code of residence. There are roughly 40, 000 zip codes in the United States--far too many variables for most machine learning algorithms to handle well.

In addition, when the number of allowable levels is very large, it becomes more likely that some of the less frequent levels may fail to show up in the training data. In our example, the researcher's training data might not include customers from some less populous zip codes, but customers from those zip codes may occur when the model is deployed. In other words, one inevitably runs into the ``novel level'' problem.

For these reasons, it is advisable to avoid these problems by converting high-cardinality categorical variables into numeric variables. There are two possible ways:

\subsubsection{Look-up codes}
Often a variable like zip code or NAICS code is really a proxy for demographic or other information of interest. For example, in a model for predicting income, the average or median income of people in a certain zip code is useful information. If one has access to external information about average income by zip code, then the zip code is simply a look-up value to that amount. Such a mapping is domain specific, and not available or appropriate in all situations.
\subsubsection{Impact coding}
Alternately, one can convert the problematic variable into a small number of numeric variables. This is known as \emph{effects coding} (\cite{Micci-Barreca}, \cite{WeightTransform}) or \emph{impact coding} (\cite{impactcoding}), or target encoding. The \pkg{ranger} random forest package (\cite{ranger}) includes an outcome-sorted ordinal effects coding (based on an idea in \cite{hastie_09_elements-of.statistical-learning} Section 9.2.4) for high-cardinality categorical variables. In \pkg{vtreat} we implement impact coding by replacing high-cardinality variables with a one-variable models for the outcome of interest (in addition to prevelance and other codes). This is best shown with an example. Here, we create a process where the input variable is a ``zip code'' that takes one of 25 possible values, and the first 3 zip codes account for 80\% of the data. The outcome is linearly related to the index of the zip code.

We start by building our example data.

\begin{Schunk}
\begin{Sinput}
R> set.seed(235)
R> Nz <- 25
R> zip <- paste0('z', format(1:Nz, justify="right"))
R> zip <- gsub(' ', '0', zip, fixed=TRUE)
R> zipval <- 1:Nz; names(zipval) <- zip
R> n <- 3; m <- Nz - n
R> p <- c(numeric(n) + (0.8/n), numeric(m) + 0.2/m)
R> N <- 1000
R> zipvar <- sample(zip, N, replace=TRUE, prob=p)
R> signal <-  zipval[zipvar] + rnorm(N)
R> d <- data.frame(zip=zipvar, 
+                  y=signal + rnorm(N))
\end{Sinput}
\end{Schunk}
We will use \pkg{vtreat} to create a \emph{treatment plan} to impact-code the zip code variable (the details will be discussed in Section~\ref{sec:basicfunctions}).
\begin{Schunk}
\begin{Sinput}
R> library("vtreat")
R> treatplan <- designTreatmentsN(d, varlist="zip", outcome="y", verbose=FALSE)
\end{Sinput}
\end{Schunk}
The treatment plan includes the observed mean of the outcome ($y$, in this case), and some information about the derived variables.

\begin{Schunk}
\begin{Sinput}
R> treatplan$meanY
\end{Sinput}
\begin{Soutput}
[1] 4.611578
\end{Soutput}
\begin{Sinput}
R> scoreFrame <- treatplan$scoreFrame
R> scoreFrame[, c('varName', 'sig', 'extraModelDegrees', 'origName', 'code')]
\end{Sinput}
\begin{Soutput}
        varName           sig extraModelDegrees origName code
1      zip_catP 2.973653e-237                24      zip catP
2      zip_catN  0.000000e+00                24      zip catN
3      zip_catD  2.675748e-01                24      zip catD
4 zip_lev_x_z01  3.495257e-28                 0      zip  lev
5 zip_lev_x_z02  4.318436e-16                 0      zip  lev
6 zip_lev_x_z03  6.960282e-08                 0      zip  lev
\end{Soutput}
\end{Schunk}

The \code{lev} variables are indicator variables that were created for the more prevalent levels; indicator variables were discussed in Section~\ref{subsec:novel}. In addition, all the levels are impact-coded into the variable \code{zip_catN}. The impact-coded variable encodes the difference between the expected outcome conditioned on zip code and the overall expected outcome: the expected ``impact'' of a particular zip code on the outcome $y$, as shown in Equation~\ref{eq:impactN}.

\begin{equation}
\label{eq:impactN}
\text{Impact}(zip) = \text{E}[y | zip] - \text{E}[y]
\end{equation}

We will not concern ourselves here with the other types of variables.

Representing levels of a categorical variable as both impact codings and indicator variables is redundant, but can be useful: indicator variables can model interactions between specific levels and other variables, while the impact coding cannot. We leave the question of which representations to use (and how) to the downstream modeling.

The function \code{vtreat::prepare} converts the original variable \code{zip} into the indicator and impact-coded variables.

\begin{Schunk}
\begin{Sinput}
R> vars <- scoreFrame$varName[!(scoreFrame$code %in% c("catP", "catD"))]
R> dtreated <- prepare(treatplan, d, pruneSig=NULL, 
+                      varRestriction=vars)
\end{Sinput}
\end{Schunk}

\begin{figure}[H]
\begin{Schunk}

\includegraphics[width=\maxwidth]{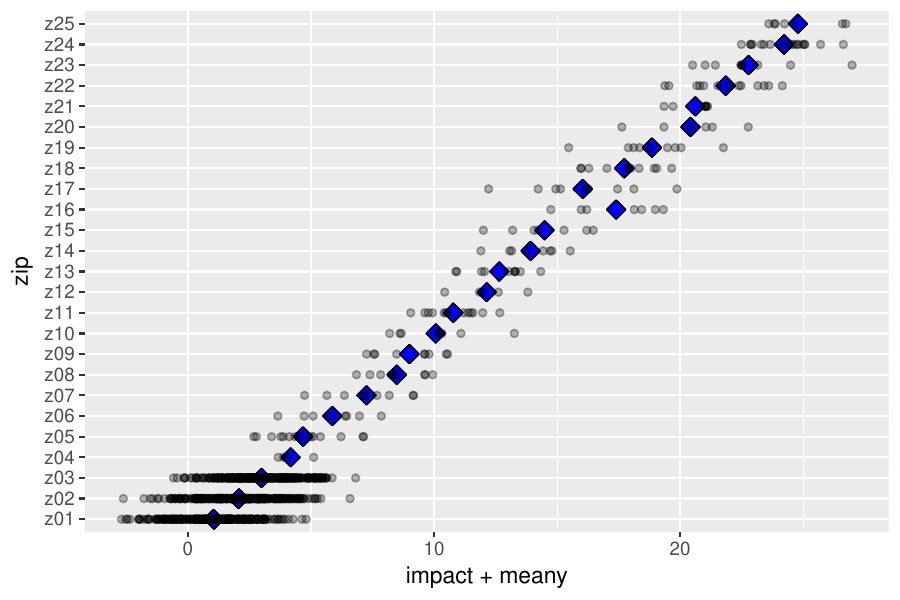} \end{Schunk}
\caption{Observed y as a function of zip, compared
to the impact-coded zip values + mean(y).}
\label{fig:zimpact}
\end{figure}

Figure~\ref{fig:zimpact} plots the impact-coded values of $zip$ added to the mean observed value $y$ (blue diamonds), compared to the observed values of $y$ for each level of zip code. We see that the impact codes successfully summarize the observed relationship between each zip code level and the outcome.

Impact coding works similarly when the outcome of interest is
categorical (two-class classification) rather than numeric. In the
case of categorical outcome $y$, with target class $target$, the
impact code represents levels of a categorical variable $x$, as shown
in Equation~\ref{eq:impactC}.

\begin{equation}
\label{eq:impactC}
\text{Impact}(x_i) = \text{logit}(\text{P}[y == target| x_i]) - \text{logit}(\text{P}[y==target])
\end{equation}

\subsubsection{Novel level impact codes}

In the example given above, Figure~\ref{fig:zimpact} shows that all the possible zip codes were present in the training data. As we noted previously, however, this may not always be true, particularly when a categorical variable takes on a great many levels with respect to the size of the training set. If these levels are encountered in the future, they are encoded as having zero impact, as shown
below.\footnote{Novel levels can instead code to a non-zero impact if
  during treatment design the pooled rare levels achieve statistical
  significance as a group.  This is similar to the novel levels
  representing uncertainty among rare levels as in
  Section~\ref{subsec:novel}.}

Here, we deliberately use a small training set, relative to the number of possible zip codes.

\begin{Schunk}
\begin{Sinput}
R> N <- 100
R> zipvar <- sample(zip, N, replace=TRUE, prob=p)
R> signal <- zipval[zipvar] + rnorm(N)
R> d <- data.frame(zip=zipvar, 
+                  y=signal+rnorm(N))
R> length(unique(d$zip))
\end{Sinput}
\begin{Soutput}
[1] 11
\end{Soutput}
\begin{Sinput}
R> omitted <- setdiff(zip, unique(d$zip))
R> print(omitted)
\end{Sinput}
\begin{Soutput}
 [1] "z04" "z05" "z08" "z09" "z12" "z17" "z18" "z19" "z20" "z21" "z22"
[12] "z23" "z24" "z25"
\end{Soutput}
\end{Schunk}

We see that not all of the possible 25 levels appear in this smaller data set. Next, 
we create a treatment plan and apply it to a data set that does contain all 25 possible levels:

\begin{Schunk}
\begin{Sinput}
R> treatplan <- designTreatmentsN(d, varlist="zip", outcome="y", verbose=FALSE)
R> dnew <- data.frame(zip = zip)
R> dtreated <- prepare(treatplan, dnew, pruneSig=NULL, 
+                     varRestriction=vars)
\end{Sinput}
\end{Schunk}

We can examine the resulting treated data frame to verify that zip codes which were missing in the training data encode to no additional impact on the outcome. This is consistent with \pkg{vtreat}'s standard novel level treatment, as discussed in Section~\ref{subsec:novel}.

\begin{Schunk}
\begin{Sinput}
R> dtreated[dnew$zip %in% omitted, "zip_catN"]
\end{Sinput}
\begin{Soutput}
 [1] 0 0 0 0 0 0 0 0 0 0 0 0 0 0
\end{Soutput}
\end{Schunk}

\subsection{Nested model bias}
\label{sec:nestedmodelbiastheory}

Care must be taken when impact coding variables -- or when using nested models in general, for example in model stacking or superlearning (\cite{superlearner}): the data used to do the impact coding should not be the same as the data used to fit the overall model.
This is because the impact coding (or the base models in superlearning) are relatively complex, high-degree-of-freedom models masquerading as low-degree-of-freedom single variables. As such, they may not be handled appropriately by downstream machine learning algorithms. In the case of impact-coded, high-cardinality categorical variables, the resulting impact coding may memorize patterns in the training data, making the variable appear more statistically significant than it really is to downstream modeling algorithms.

This is best shown with an example. Consider the following data frame. The outcome (binary classification) only depends on the ``good'' variables, not on the (also high degree of freedom) ``bad'' variables. Modeling such a data set runs a high risk of overfit.

\begin{Schunk}
\begin{Sinput}
R> set.seed(2262)
R> nLev <- 500
R> n <- 3000
R> d <- data.frame(xBad1=sample(paste('level', 1:nLev, sep=''), n, replace=TRUE), 
+                  xBad2=sample(paste('level', 1:nLev, sep=''), n, replace=TRUE), 
+                  xGood1=sample(paste('level', 1:nLev, sep=''), n, replace=TRUE), 
+                  xGood2=sample(paste('level', 1:nLev, sep=''), n, replace=TRUE))
R> d$y <- (0.2*rnorm(nrow(d)) + 0.5*ifelse(as.numeric(d$xGood1)>nLev/2, 1, -1) +
+          0.3*ifelse(as.numeric(d$xGood2)>nLev/2, 1, -1))>0
R> d$rgroup <- sample(c("cal", "train", "test"), nrow(d), replace=TRUE, 
+                     prob=c(0.6, 0.2, 0.2))
R> 
R> plotRes <- function(d, predName, yName, title) {
+    print(title)
+    tab <- table(truth=d[[yName]], pred=d[[predName]]>0.5)
+    print(tab)
+    diag <- sum(vapply(seq_len(min(dim(tab))), 
+                       function(i) tab[i, i], numeric(1)))
+    acc <- diag/sum(tab)
+    # depends on both truth and target being logicals
+    # and FALSE ordered before TRUE
+    sens <- tab[2, 2]/sum(tab[2, ])
+    spec <- tab[1, 1]/sum(tab[1, ])
+    print(paste('accuracy', format(acc, scientific=FALSE, digits=3)))
+    print(paste('sensitivity', format(sens, scientific=FALSE, digits=3)))
+    print(paste('specificity', format(spec, scientific=FALSE, digits=3)))
+  }
\end{Sinput}
\end{Schunk}

\subsubsection{The wrong way: naive data partitioning}
First, we will partition the data into a training set and a holdout set, and create the \pkg{vtreat} treatment plan. This plan will include impact codings for the high-cardinality categorical variables \code{xBad}$i$.

\begin{Schunk}
\begin{Sinput}
R> dTrain <- d[d$rgroup!='test', , drop=FALSE]
R> dTest <- d[d$rgroup=='test', , drop=FALSE]
R> treatments <- vtreat::designTreatmentsC(dTrain, 
+                       varlist = c('xBad1', 'xBad2', 'xGood1', 'xGood2'), 
+                       outcomename='y', outcometarget=TRUE, 
+                       verbose=FALSE)
R> dTrainTreated <- vtreat::prepare(treatments, dTrain, pruneSig=NULL)
\end{Sinput}
\end{Schunk}

Next we fit a model to the treated training data. We will fit a logistic regression model, but the effects shown are possible with any other modeling algorithm.

\begin{Schunk}
\begin{Sinput}
R> m1 <- glm(y~xBad1_catB + xBad2_catB + xGood1_catB + xGood2_catB, 
+            data=dTrainTreated, family=binomial(link='logit'))
R> print(summary(m1))
\end{Sinput}
\begin{Soutput}

Call:
glm(formula = y ~ xBad1_catB + xBad2_catB + xGood1_catB + xGood2_catB, 
    family = binomial(link = "logit"), data = dTrainTreated)

Deviance Residuals: 
     Min        1Q    Median        3Q       Max  
-2.75305  -0.00262   0.00000   0.00294   2.60808  

Coefficients:
            Estimate Std. Error z value Pr(>|z|)    
(Intercept) 0.001452   0.124260   0.012    0.991    
xBad1_catB  0.996016   0.160774   6.195 5.82e-10 ***
xBad2_catB  1.233834   0.193522   6.376 1.82e-10 ***
xGood1_catB 1.067978   0.095673  11.163  < 2e-16 ***
xGood2_catB 1.485611   0.187671   7.916 2.45e-15 ***
---
Signif. codes:  0 '***' 0.001 '**' 0.01 '*' 0.05 '.' 0.1 ' ' 1

(Dispersion parameter for binomial family taken to be 1)

    Null deviance: 3351.98  on 2417  degrees of freedom
Residual deviance:  419.19  on 2413  degrees of freedom
AIC: 429.19

Number of Fisher Scoring iterations: 10
\end{Soutput}
\end{Schunk}

Note the low residual deviance of the model, and that the ``bad'' variables appear significant in the model. For the classification task, we use a model score of 0.5 as the threshold between positive and negative classes. Classification performance on the training set appears quite good.

\begin{Schunk}
\begin{Sinput}
R> dTrain$predM1 <- predict(m1, newdata=dTrainTreated, type='response')
R> plotRes(dTrain, 'predM1', 'y', 'model1 on train')
\end{Sinput}
\begin{Soutput}
[1] "model1 on train"
       pred
truth   FALSE TRUE
  FALSE  1158   44
  TRUE     45 1171
[1] "accuracy 0.963"
[1] "sensitivity 0.963"
[1] "specificity 0.963"
\end{Soutput}
\end{Schunk}

However, the model does not perform nearly as well on the holdout set -- a clear case of overfit.

\begin{Schunk}
\begin{Sinput}
R> dTestTreated <- vtreat::prepare(treatments, dTest, pruneSig=NULL)
R> dTest$predM1 <- predict(m1, newdata=dTestTreated, type='response')
R> plotRes(dTest, 'predM1', 'y', 'model1 on test')
\end{Sinput}
\begin{Soutput}
[1] "model1 on test"
       pred
truth   FALSE TRUE
  FALSE   208   63
  TRUE     81  230
[1] "accuracy 0.753"
[1] "sensitivity 0.74"
[1] "specificity 0.768"
\end{Soutput}
\end{Schunk}

One way to defend against this would have been to examine the variable significance estimates
in \code{treatments$scoreFrame} which are estimated out of sample. We discuss variable significance in Section~\ref{sec:varchoiceheuristic}.
Here, however, we wish to discuss even stronger techniques: calibration sets and simulated out of sample cross-frames.

\subsubsection{The right way: a calibration set}
\label{sec:useofcalset}

Consider any trained statistical model (in this case the design and subsequent application of a treatment plan) as a two-argument function
$f(A, B)$. The first argument is the training data and the second argument is the application data. Using the \pkg{magrittr} (\cite{magrittr}) pipe notation, we can write $f(A, B)$ as
\begin{quote}
\code{designTreatmentsC(A) \%>\% prepare(B)}, 
\end{quote}
which produces a treated data frame.

When we use the same data in both places to build our training frame, as in
\[
TrainTreated = f(TrainData, TrainData), 
\]
we are not doing a good job simulating the future application of $f(, )$, which will be 
\[
f(TrainData, FutureData)
\]

To improve the quality of our simulation we can call
\[
TrainTreated = f(CalibrationData, TrainData)
\]
where $CalibrationData$ and $TrainData$ are disjoint datasets. We expect this to be a good imitation of future $f(CalibrationData, FutureData)$.

To see this, we now try the same problem as above, partitioning the
data into training, calibration, and holdout sets. The impact coding
is fit to the calibration set, and the overall model is fit to the
training set.

\begin{Schunk}
\begin{Sinput}
R> dCal <- d[d$rgroup=='cal', , drop=FALSE]
R> dTrain <- d[d$rgroup=='train', , drop=FALSE]
R> dTest <- d[d$rgroup=='test', , drop=FALSE]
R> treatments <- vtreat::designTreatmentsC(dCal, 
+                  varlist = c('xBad1', 'xBad2', 'xGood1', 'xGood2'), 
+                  outcomename='y', outcometarget=TRUE, 
+                  verbose=FALSE)
R> dTrainTreated <- vtreat::prepare(treatments, dTrain, 
+                          pruneSig=NULL)
R> newvars <- setdiff(colnames(dTrainTreated), 'y')
R> m1 <- glm(y~xBad1_catB + xBad2_catB + xGood1_catB + xGood2_catB, 
+            data=dTrainTreated, family=binomial(link='logit'))
R> dTrain$predM1 <- predict(m1, newdata=dTrainTreated, type='response')
R> print(summary(m1))
\end{Sinput}
\begin{Soutput}

Call:
glm(formula = y ~ xBad1_catB + xBad2_catB + xGood1_catB + xGood2_catB, 
    family = binomial(link = "logit"), data = dTrainTreated)

Deviance Residuals: 
    Min       1Q   Median       3Q      Max  
-2.3490  -0.3823   0.3055   0.4139   2.4888  

Coefficients:
            Estimate Std. Error z value Pr(>|z|)    
(Intercept) -0.08133    0.12605  -0.645    0.519    
xBad1_catB  -0.01950    0.02518  -0.774    0.439    
xBad2_catB  -0.01866    0.02610  -0.715    0.475    
xGood1_catB  0.25711    0.01964  13.092   <2e-16 ***
xGood2_catB  0.03668    0.02407   1.524    0.127    
---
Signif. codes:  0 '***' 0.001 '**' 0.01 '*' 0.05 '.' 0.1 ' ' 1

(Dispersion parameter for binomial family taken to be 1)

    Null deviance: 768.87  on 554  degrees of freedom
Residual deviance: 412.40  on 550  degrees of freedom
AIC: 422.4

Number of Fisher Scoring iterations: 5
\end{Soutput}
\end{Schunk}

Note that this model successfully recognizes that the \code{xBad}$i$ variables are not significant. Classification performance on the training set is good.

\begin{Schunk}
\begin{Sinput}
R> plotRes(dTrain, 'predM1', 'y', 'model1 on train')
\end{Sinput}
\begin{Soutput}
[1] "model1 on train"
       pred
truth   FALSE TRUE
  FALSE   238   31
  TRUE     40  246
[1] "accuracy 0.872"
[1] "sensitivity 0.86"
[1] "specificity 0.885"
\end{Soutput}
\end{Schunk}

Classification performance on the holdout set is now similar to training. The three way split of the data has resolved the overfit issue in two ways: training performance is closer to test performance, and test performance is better than that with the model fit using the naive data partition.

\begin{Schunk}
\begin{Sinput}
R> dTestTreated <- vtreat::prepare(treatments, dTest, 
+                                  pruneSig=NULL)
R> dTest$predM1 <- predict(m1, newdata=dTestTreated, type='response')
R> plotRes(dTest, 'predM1', 'y', 'model1 on test')
\end{Sinput}
\begin{Soutput}
[1] "model1 on test"
       pred
truth   FALSE TRUE
  FALSE   241   30
  TRUE     48  263
[1] "accuracy 0.866"
[1] "sensitivity 0.846"
[1] "specificity 0.889"
\end{Soutput}
\end{Schunk}

\subsubsection[Another right way: cross-validation and vtreat]{Another right way: cross-validation and \pkg{vtreat}}
\label{sec:nestedmodeling}
Returning to our $f(A, B)$ notation, another, more statistically efficient approach is to build a cross validated version of $f$. We split $TrainData$ into a list of 3 disjoint row intervals: $Train1, Train2, Train3$. Instead of computing $f(TrainData, TrainData)$ compute:
\begin{eqnarray*}
TrainTreated &=& f(Train2+Train3, Train1) + \\
             & & f(Train1+Train3, Train2) + \\
             & &  f(Train1+Train2, Train3)
\end{eqnarray*}
where `+' denotes \code{rbind()}.

This looks a lot like $f(TrainData, TrainData)$ except it has the important property that no row in the right-hand side is ever worked on by a model built using that row -- a key characteristic that future data will have. We therefore have a good imitation of
$f(TrainData, FutureData)$.

In other words, we use cross validation ideas to simulate future data. The key point is that we can apply cross validation to any two argument function $f(A, B)$ and not only to functions of the form $f(A, B)$ = \code{buildModel(A) \%>\% scoreData(B)}. We can use this formulation in stacking or super-learning with $f(A, B)$ of the form \code{buildSubModels(A) \%>\% combineModels(B)}; the idea applies to improving ensemble methods in general.

\begin{figure}[H]
\includegraphics[width=5in]{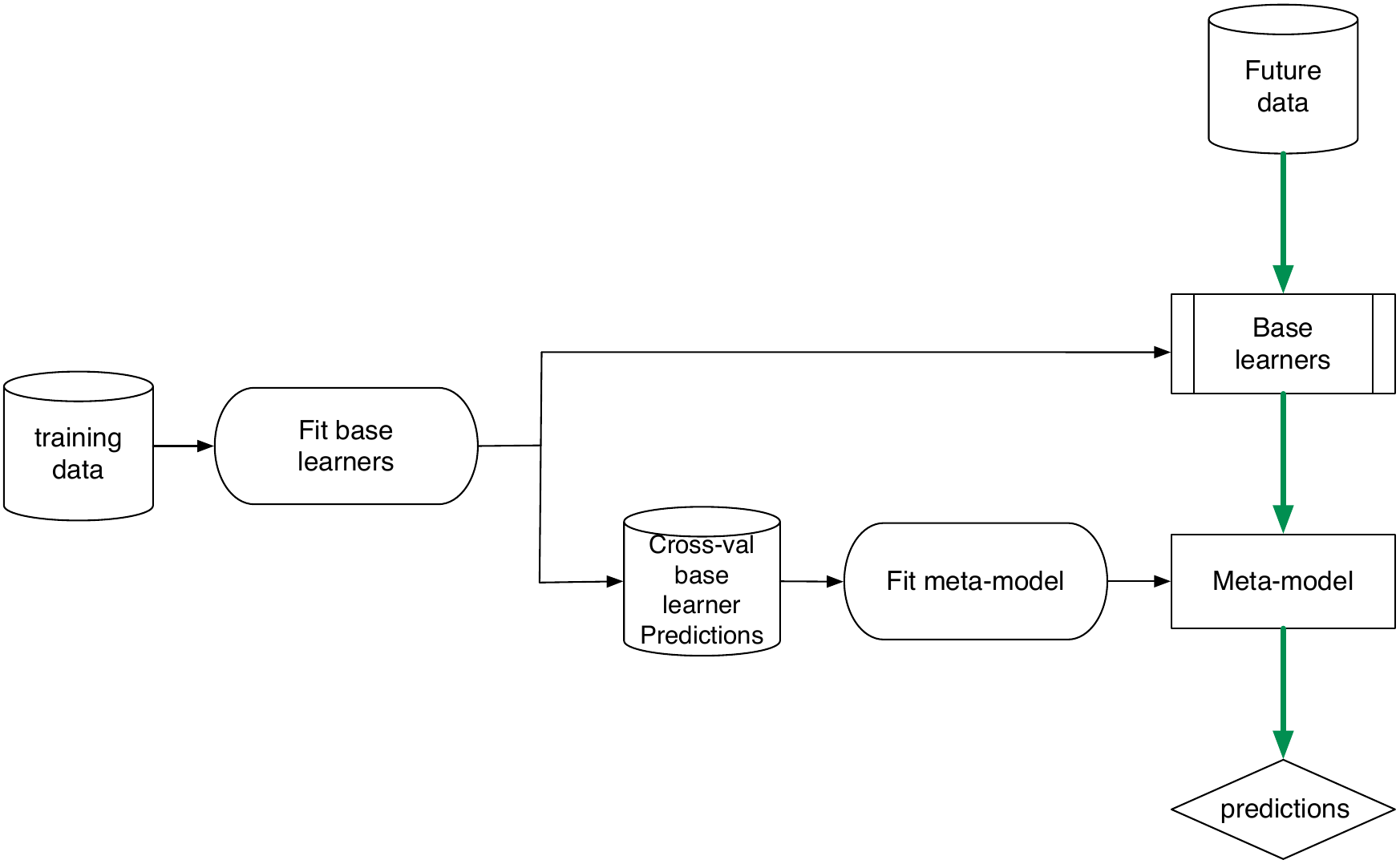}
\caption{Schematic of stacking: the meta-model is fit using cross-validated base learner predictions.}
\label{fig:supermodel}
\end{figure}

See \cite{MitGee09}, \cite{DBLP:journals/sigkdd/PerlichS10}, \cite{MitGee09}, and \cite{superlearner} for further discussion of cross-validating submodels, or base learners, in the context of stacked models, or superlearning. In super learning cross validation techniques are used to simulate having built base learner predictions on novel data. The simulated out-of-sample applications of these base learners (and not the base learners themselves) are then used as input data for the next stage learner, or meta-model. In future application the actual base learners are applied and their immediate outputs are used by the meta-model. This is shown in
Figure~\ref{fig:supermodel}.

\begin{figure}[H]
\includegraphics[width=5in]{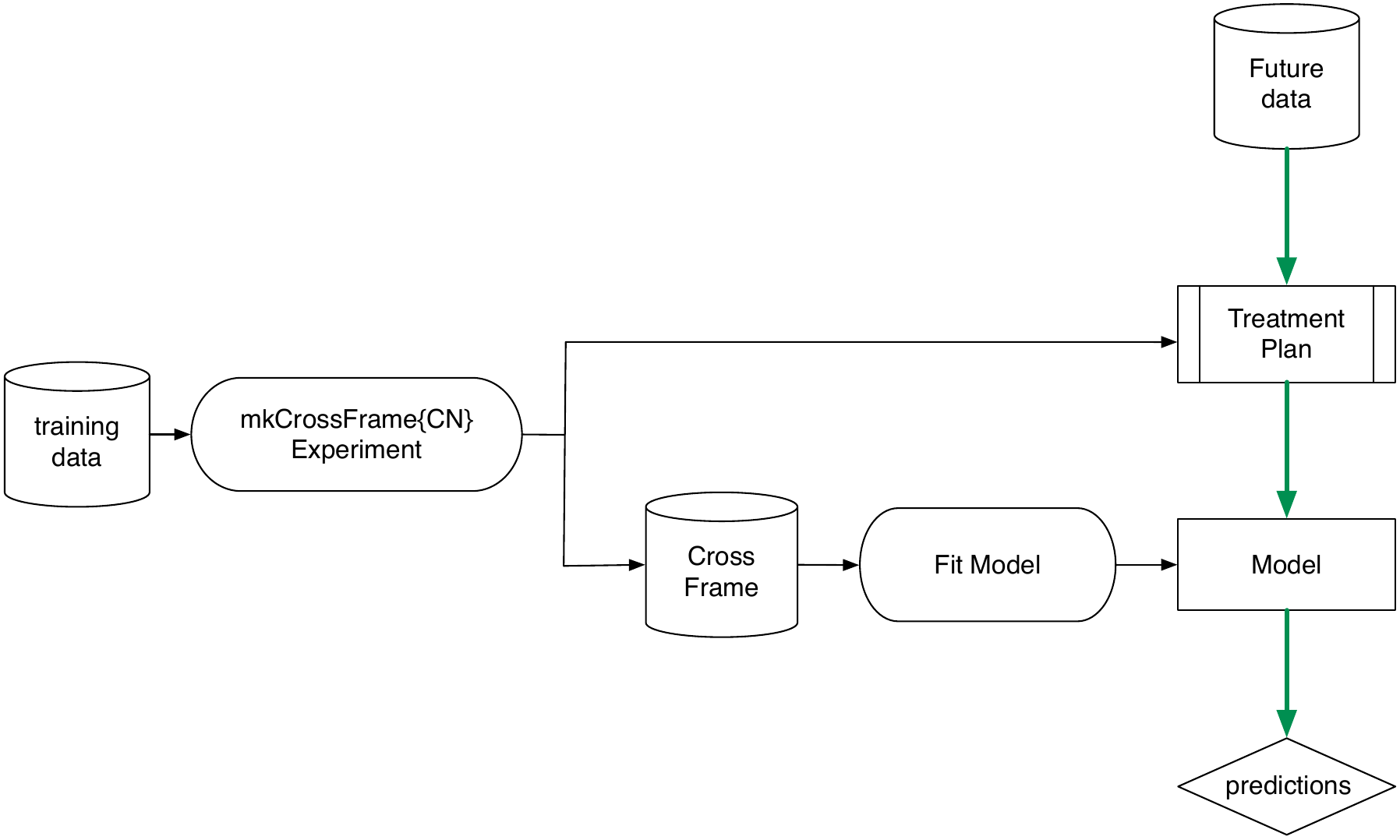}
\caption{Schematic of a model fit using a vtreat crossframe.}
\label{fig:crossframes}
\end{figure}

In \pkg{vtreat} the ``base learners'' are single variable treatments and the outer model construction is left to the practitioner, using what
we refer to as \emph{crossframes} for simulation, rather than preparing the training set using a treatment plan. In application the treatment plan is used. This is shown in Figure~\ref{fig:crossframes}.

Note that Figure~\ref{fig:supermodel} and Figure~\ref{fig:crossframes} show the identical structure. In fact (though it was developed independently) one can think of \pkg{vtreat} as a stacked model. We will discuss crossframes in Section~\ref{sec:crossframes}.

\subsection{Wide data: variable significance and variable pruning}
\label{sec:varchoiceheuristic}

Wide data sets--data with many variables relative to the number of exemplars--are computationally difficult for some modeling procedures; and more importantly, they can lead to overfit models that generalize poorly on new data. In extreme cases, wide data can fool modeling procedures into finding models that look good on training data, even when that data has no signal (\cite{screening}). Too many irrelevant variables can also appreciably slow down model fitting. For these reasons, it may be advisable to prune irrelevant variables before modeling.

Standard approaches to variable pruning include stepwise regression (\cite{Faraway}), L1 (lasso) regularization (\cite{hastie_09_elements-of.statistical-learning}), and the use of Random Forest variable importance estimates (\cite{Genuer}). Stepwise regression in particular suffers from a multiple-experiment bias, and on the bias caused by repeated evaluation of interim models on the same data set.

As an alternative to these approaches, \pkg{vtreat} offers estimates of variable significance and the option of pruning variables based on these significances during the data preparation step. The operational details will be discussed in Section~\ref{sec:sigop}. Variable significances are based on the significance of the corresponding single-variable model. For problems with a numeric outcome, the significance is based on the F statistic of a single variable linear regression; for problems with a categorical outcome, the significance is based on the $\chi^2$ statistic of a single variable logistic regression.

Care must be taken when estimating the significance of categorical variables; recall that a categorical variable with $k$ levels is equivalent to $k-1$ indicator variables. These additional degrees of freedom must be accounted for when estimating the significance of the F or $\chi^2$ statistic.

Variable pruning based on these significances is of course only heuristic. The primary assumption of this heuristic is that a useful variable has a signal that could be detected by a small linear or logistic model \textit{even} if the original relation is complex or non-linear. This can miss variables that are in fact useful in a larger joint model, but which happen to look orthogonal to the outcome when taken alone.\footnote{This negative effect is typically associated with interactions, but can also be an undesirable feature even of models that are linear over the original variables.  Familiar examples include Simpson's paradox and model coefficients that change sign upon introduction of additional variables.}

\subsubsection{Choosing the significance threshold}

We can interpret the significance of a variable as the probability that a non-signaling variable would have an F or $\chi^2$ statistic as large as the value observed. If we have 100 non-signaling variables and we are accepting variables with a significance value less than $p$, then we expect to erroneously accept about $100 p$ of the non-signaling variables. So the significance threshold for variable pruning is the false positive rate that we are willing to accept. This false positive rate should be greater than zero, as modeling algorithms should be able to tolerate a few irrelevant variables. As a rule of thumb, we recommend setting the pruning threshold to $p = 1/n_{var}$, where $n_{var}$ is the number of candidate variables.

\subsubsection{A worked example}

As an example, we give a regression problem with two numeric inputs (one signal \code{sigN}, one noise, \code{noiseN}) and two high-cardinality categoric inputs (one signal, \code{sigC}, and one noise, \code{noiseC}, both with 100 levels).

\begin{Schunk}
\begin{Sinput}
R> set.seed(22451)
R> N <- 500
R> sigN <- rnorm(N)
R> noiseN <- rnorm(N)
R> Nlevels <- 100
R> zip <- paste0('z', format(1:Nlevels, justify="right"))
R> zip <- gsub(' ', '0', zip, fixed=TRUE)
R> 
R> zipval <- runif(Nlevels); names(zipval)=zip
R> sigC <- sample(zip, size=N, replace=TRUE)
R> noiseC <- sample(zip, size=N, replace=TRUE)
R> 
R> y <- sigN + zipval[sigC] + rnorm(N)
R> df <- data.frame(sN = sigN, nN=noiseN, 
+                  sC = sigC, nC=noiseC, y=y)
\end{Sinput}
\end{Schunk}

Designing a treatment plan from this data gives us the following derived variable types (we ignore the \code{catP} and \code{catD} variable types).

\begin{Schunk}
\begin{Sinput}
R> library("vtreat")
R> treatplan <- designTreatmentsN(df, 
+                               varlist=setdiff(colnames(df), "y"), 
+                               outcomename="y", 
+                               verbose=FALSE)
R> sframe <- treatplan$scoreFrame
R> vars <- sframe$varName[!(sframe$code %in% c("catP", "catD"))]
R> sframe[sframe$varName %in% vars, 
+         c("varName", "sig", "extraModelDegrees")]
\end{Sinput}
\begin{Soutput}
         varName          sig extraModelDegrees
1             sN 1.592457e-71                 0
2             nN 1.369134e-01                 0
4        sC_catN 4.309992e-01                99
7        nC_catN 3.228694e-02                99
9  nC_lev_x_z015 6.141466e-01                 0
10 nC_lev_x_z023 6.543513e-01                 0
11 nC_lev_x_z030 6.523778e-01                 0
12 nC_lev_x_z065 2.140222e-01                 0
13 nC_lev_x_z068 4.248945e-01                 0
14 nC_lev_x_z084 7.308342e-01                 0
\end{Soutput}
\end{Schunk}

For each derived variable, \pkg{vtreat} reports a significance estimate and any extra degrees of freedom in the corresponding ``one variable model'', which helps the user reproduce the corresponding significance calculation. For a categorical variable, the extra degrees of freedom are the number of observed levels minus one. We can plot the significance estimates:

\begin{figure}[H]
\begin{Schunk}

\includegraphics[width=\maxwidth]{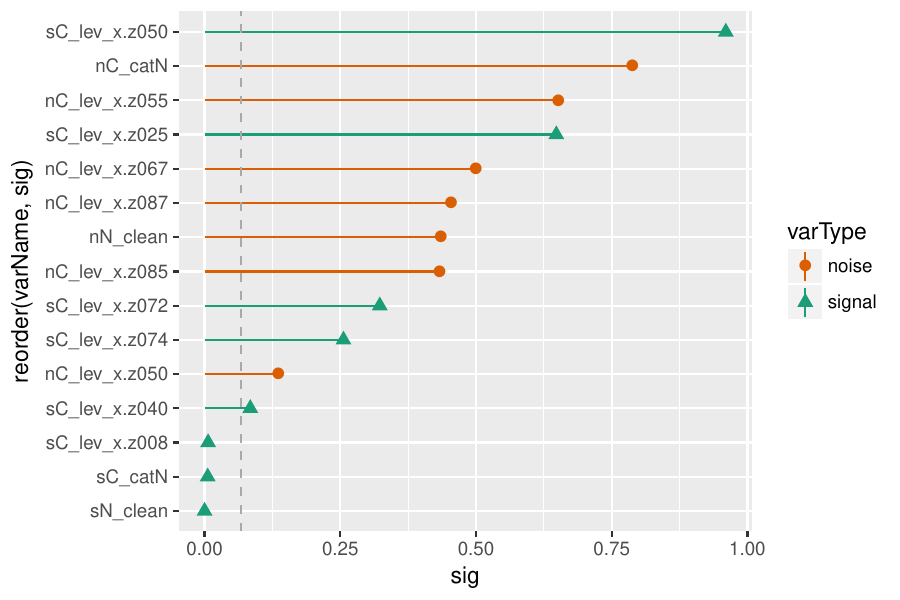} \end{Schunk}
\caption{Estimated significances of derived variables.}
\label{fig:sigstem}
\end{figure}

The dashed line in Figure~\ref{fig:sigstem} shows the proposed pruning threshold of $1/n_{var}$, 
where being left of the threshold is good.  This accepts the signaling
numeric variable \code{sN}, the impact coding of the signaling
categorical variable \code{sC_catN}, and an indicator corresponding to
one of \code{sC}'s levels. All derived variables corresponding to the
original noise variables are rejected. Note that several signaling
indicator variables are also rejected. This is because this data set
is not large enough for these variables to achieve
significance. However, much of their utility is still captured in the
\code{sc_catN} impact variable.

The function \code{vtreat::prepare} takes the argument \code{pruneSig} to pass in the desired pruning threshold.

\begin{Schunk}
\begin{Sinput}
R> pruneSig <- 1/length(vars)
R> dfTreat <- prepare(treatplan, df, pruneSig=pruneSig, 
+                    varRestriction=vars)
R> head(dfTreat)
\end{Sinput}
\begin{Soutput}
          sN    nC_catN          y
1  0.2035616  0.3277152  0.8354587
2 -0.3156755 -0.2389059 -1.2045723
3 -0.5220015 -0.5472413  1.8379891
4  0.2917703 -0.2134639  1.2366201
5 -0.2233112  0.9132010  1.4587825
6 -0.1796532  0.6846848  0.3502096
\end{Soutput}
\end{Schunk}

\section[Basic functions of vtreat]{Basic functions of \pkg{vtreat}}
\label{sec:basicfunctions}

In this section we will discuss the operational aspects of using the
\pkg{vtreat} package, and explain some of the components and
conventions of the package.  All \pkg{vtreat} methods have
detailed matching \code{help()} and vignettes\footnote{\url{http://winvector.github.io/vtreat/}}
with examples.
The purpose of this section is to document how to use
the \pkg{vtreat} package, leaving justification to Section~\ref{sec:theory}.

\pkg{vtreat} is designed to prepare data for predictive modeling where
the quantity to be predicted is either numeric or is treated as a
binomial classification target.  \pkg{vtreat} can also prepare data
where there is no quantity to be predicted, but this is not its
primary purpose.

The basic data preparation and use process is:
\begin{enumerate}
\item Use \code{vtreat::designTreatments*} to collect statistics on a
\code{data.frame} and produce a \textit{treatment plan}.
\item Use the treatment plan to process subsequent \code{data.frame}s for model training
and model application, via the function \code{vtreat::prepare}.
\end{enumerate}

\begin{figure}[H]
\includegraphics[width=5in]{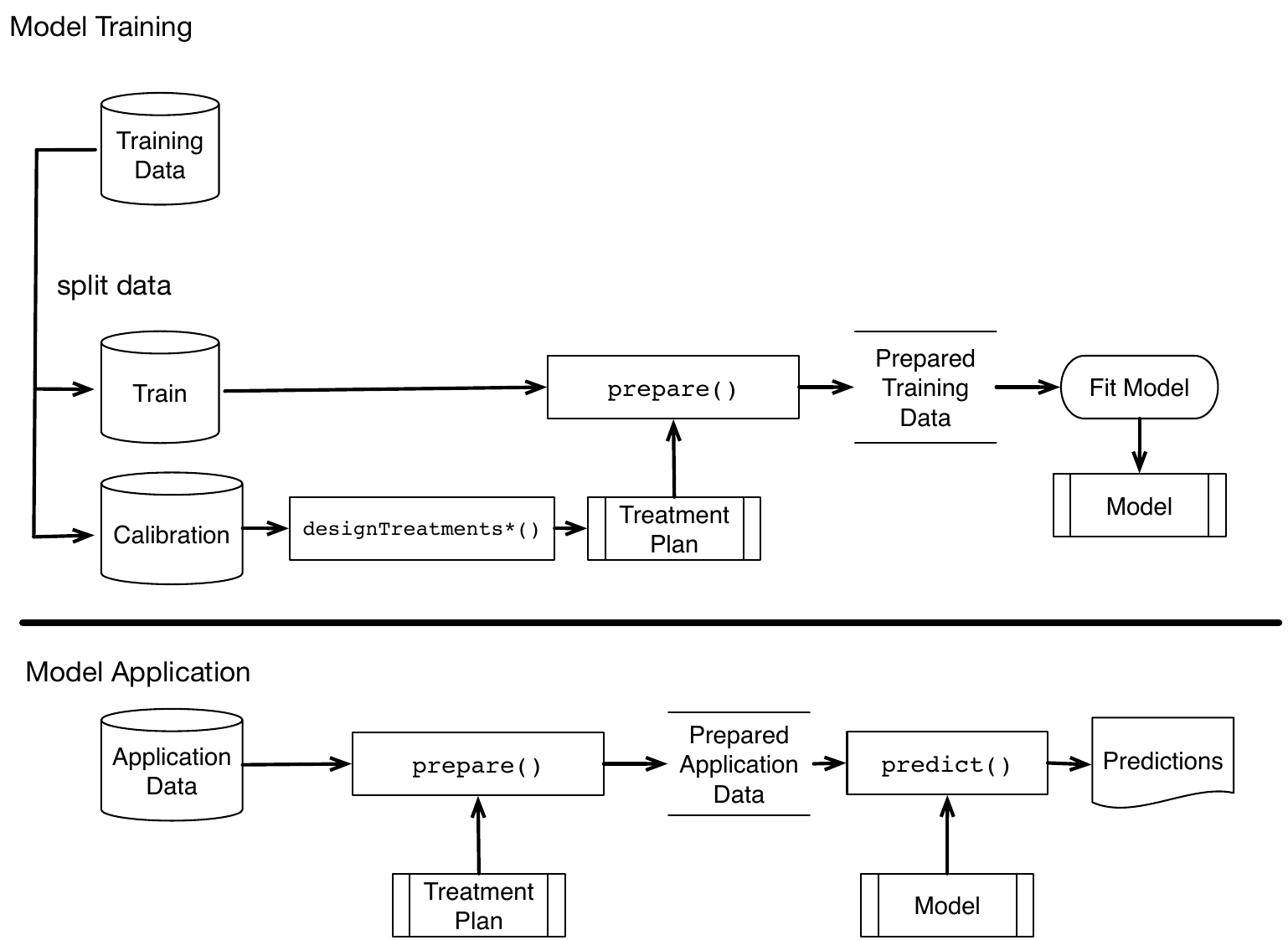}
\caption{Schematic of vtreat data preparation and use process.}
\label{fig:vtreatProcess}
\end{figure}

A processed \code{data.frame} has only numeric columns (other than
the outcome column\footnote{Note: during treatment design it is expected that the outcome
  column itself does not have any \code{Infinite}/\code{NA}/\code{NaN}
  values and takes on more than one value.}), and has no
\code{Infinite}/\code{NA}/\code{NaN} values in the derived variable
columns.
\pkg{vtreat} serves as a powerful alternative to
\code{model.matrix}, which is implicitly used in many \proglang{R} modeling tasks.

As discussed in Section~\ref{sec:nestedmodelbiastheory}, when the data preparation includes
the impact coding of high-cardinality categorical variables, one should create the treatment plan on data
distinct from the data used to train a model. When using the \code{designTreatments*}/\code{prepare} pattern we recommend
using a three-way data partition where data is organized into a
calibration set (used for the \pkg{vtreat} design phase), a training set
(used for subsequent predictive model construction), and a test set
(used for final model evaluation).  This is shown in Figure~\ref{fig:vtreatProcess}, and we
include a worked example of this technique in Section~\ref{sec:nestedmodelbiastheory}.

Alternatively, one can use a \code{mkCrossFrame*Experiment}/\code{prepare} pattern with
a more statistically efficient two way partition where training data is used for the
design of variable treatments and modeling, and the test data is again
used for model evaluation. We discuss this work pattern in Section~\ref{sec:crossframes};
we showed the process schematically in Figure~\ref{fig:crossframes}.

We will demonstrate \pkg{vtreat} operations using the following artificial \code{data.frame}, which manifests the data issues that \pkg{vtreat} mitigates.

\begin{Schunk}
\begin{Sinput}
R> d <- data.frame(
+     x=c('a', 'a', 'b', 'b', NA), 
+     z=c(0, 1, 2, NA, 4), 
+     y=c(TRUE, TRUE, FALSE, TRUE, TRUE), 
+     stringsAsFactors = FALSE)
R> d$yN <- as.numeric(d$y)
R> print(d)
\end{Sinput}
\begin{Soutput}
     x  z     y yN
1    a  0  TRUE  1
2    a  1  TRUE  1
3    b  2 FALSE  0
4    b NA  TRUE  1
5 <NA>  4  TRUE  1
\end{Soutput}
\end{Schunk}

Using \pkg{vtreat} we can process this \code{data.frame} a number of
ways.  In all cases we get back a treatment plan object (itself a
portion of a larger structure in the case of
\code{mkCrossFrame*Experiment}) and on this treatments object there is
a new \code{data.frame} value called \code{scoreFrame} that documents
the variable transformations.  These steps and data structures are demonstrated below.

\subsection{Designing and applying a treatment plan for a numeric target}

In this section, we will create a treatment plan from the data frame \code{d} in preparation for fitting a model to predict the numeric target \code{yN}.

\begin{Schunk}
\begin{Sinput}
R> library("vtreat")
R> treatments <- designTreatmentsN(d, c('x', 'z'), 'yN')
\end{Sinput}
\end{Schunk}
\begin{Schunk}
\begin{Sinput}
R> scols <- c('varName', 'sig', 'extraModelDegrees', 'origName', 'code')
R> print(treatments$scoreFrame[, scols])
\end{Sinput}
\begin{Soutput}
    varName       sig extraModelDegrees origName  code
1    x_catP 0.6850376                 2        x  catP
2    x_catN 0.5815357                 2        x  catN
3    x_catD 0.4950253                 2        x  catD
4         z 0.8798694                 0        z clean
5   z_isBAD 0.6850376                 0        z isBAD
6  x_lev_NA 0.6850376                 0        x   lev
7 x_lev_x_a 0.4950253                 0        x   lev
8 x_lev_x_b 0.2722284                 0        x   lev
\end{Soutput}
\end{Schunk}

Each new column or variable produced by \pkg{vtreat} is represented in the \pkg{scoreFrame} as
a row named by the column \code{varName}.  Reported information about the new variables includes:

\begin{itemize}
\item \code{sig} The significance of a single variable linear model built using the variable to
predict the target outcome.  The significance is based on the appropriate F-test.
\item \code{extraModelDegrees} This is how many extra degrees of freedom the new variable represents.  Notice
this value is zero for most variables, and it is the number of levels minus one for derived columns that represent
re-encodings of entire ranges of categorical variables.
\item \code{origName} The original variable that this new variable is derived from.
\item \code{code} The type of transform used to produce the derived variable, also called the \pkg{vtreat} variable type.
\end{itemize}

For \code{designTreatmentsN} the possible derived variable types are:

\begin{itemize}
\item \code{clean}: a numeric variable with all
  NA/NaN/infinite values replaced by the mean value of
  the non-NA/NaN/infinite examples of the variable.
\item \code{is\_Bad} : a companion to the \code{clean} treatment.  \code{is\_Bad} is an indicator that indicates a value replacement has occurred.  For many noisy data-sets this column can be more informative than the clean column!
\item \code{lev} : a 0/1 indicator indicating a particular value of a categorical variable was present.  For example \code{x_lev_x.a} is 1 when the original \code{x} variable had a value of ``a''.  These indicators are essentially variables representing explicit encoding of levels as dummy variables.  In some cases a special level code is used to represent pooled rare values.
\item \code{cat\_N} : a single variable regression model of the difference in outcome expectation conditioned on the observed value of the original variable.  In our example: \code{x\_catN} = $\text{E}[yN|x] - \text{E}[yN]$.  This encoding is especially useful for categorical variables that have a large number of levels, but be aware it can obscure degrees of freedom if not used properly.
\item \code{cat\_P} : the prevalence (frequency) of each categorical level in the training data.  This indicates if the original level was rare or common. Not always directly useful in the model, but can be useful in interactions.
\item \code{cat\_D} : the within-group deviation of the outcome conditioned on each categorical level in the training data. This indicates whether the outcome value is concentrated or diffuse with respect to a particular level. Not always directly useful in the model, but can be useful in interactions.
\end{itemize}

Once we have \code{treatments} we can use it to \code{prepare} or transform any data frame that has at least the set of input columns we designed on.  Below we demonstrate the procedure on our simple \code{data.frame}. The \code{pruneSig} argument is the mandatory user supplied significance pruning level; setting it to \code{NULL} turns off pruning.

\begin{Schunk}
\begin{Sinput}
R> dTreated <- prepare(treatments, d, pruneSig=NULL)
R> print(dTreated)
\end{Sinput}
\begin{Soutput}
  x_catP x_catN    x_catD    z z_isBAD x_lev_NA x_lev_x_a x_lev_x_b yN
1    0.4    0.2 0.0000000 0.00       0        0         1         0  1
2    0.4    0.2 0.0000000 1.00       0        0         1         0  1
3    0.4   -0.3 0.7071068 2.00       0        0         0         1  0
4    0.4   -0.3 0.7071068 1.75       1        0         0         1  1
5    0.2    0.2 0.7071068 4.00       0        1         0         0  1
\end{Soutput}
\end{Schunk}

If present in the input the outcome column is copied into the prepared \code{data.frame}. The resulting data frame
\code{dTreated} can now be used to safely fit a model to predict the outcome \code{yN}.

\subsection{Designing and applying a treatment plan for a categorical target}

Preparing a treatment plan for a binomial classification problem is similar to the preparation for a numeric or regression problem.
The difference is we call \code{designTreatmentsC} and need to supply which value of the target variable is considered to be the
positive or target category.  In this case the outcome variable is \code{y}, and we will use the value \code{TRUE} as our target.

\begin{Schunk}
\begin{Sinput}
R> treatments <- designTreatmentsC(d, c('x', 'z'), 'y', TRUE)
\end{Sinput}
\end{Schunk}
\begin{Schunk}
\begin{Sinput}
R> print(treatments$scoreFrame[, scols])
\end{Sinput}
\begin{Soutput}
    varName       sig extraModelDegrees origName  code
1    x_catP 0.4771618                 2        x  catP
2    x_catB 0.5445007                 2        x  catB
3         z 0.8341162                 0        z clean
4   z_isBAD 0.4771618                 0        z isBAD
5  x_lev_NA 0.4771618                 0        x   lev
6 x_lev_x_a 0.2763528                 0        x   lev
7 x_lev_x_b 0.1352282                 0        x   lev
\end{Soutput}
\end{Schunk}

The \code{scoreFrame} is much the same, except that the significance reports the quality of a single variable logistic regression model, so it uses the $\chi^2$ test.
For categorical targets the possible derived variable types are as follows:

\begin{itemize}
\item \code{clean} : a numeric variable passed through with all NA/NaN/infinite values replaced with either zero or mean value of the non-NA/NaN/infinite examples of the variable.
\item \code{is\_Bad} : a companion to the \code{clean} treatment.  \code{is\_Bad} is an indicator that indicates a value replacement has occurred.  For many noisy datasets this column can be more informative than the clean column!
\item \code{lev} : a 0/1 indicator indicating a particular value of a categorical variable was present.  For example \code{x_lev_x.a} is 1 when the original \code{x} variable had a value of ``a''.  These indicators are essentially variables representing explicit encoding of levels as dummy variables. In some cases a special level code is used to represent pooled rare values.
\item \code{cat\_B} : a single variable Bayesian model of the difference in logit-odds in outcome from the mean distribution, conditioned on the observed value of the original variable.  In our example: \code{x\_catB} = $\text{logit}(\text{P}[y==target|x]) - \text{logit}(\text{P}[y==target])$.  This encoding is especially useful for categorical variables that have a large number of levels, but be aware it can obscure degrees of freedom if not used properly.
\item \code{cat\_P} : the prevalence (frequency) of each categorical level in the training data.  This indicates if the original level was rare or common. Not always directly useful in the model, but can be useful in interactions.
\end{itemize}

And we can prepare \code{data.frame}s as before.

\begin{Schunk}
\begin{Sinput}
R> dTreated <- prepare(treatments, d, pruneSig=NULL)
R> print(dTreated)
\end{Sinput}
\begin{Soutput}
  x_catP    x_catB    z z_isBAD x_lev_NA x_lev_x_a x_lev_x_b     y
1    0.4  8.517318 0.00       0        0         1         0  TRUE
2    0.4  8.517318 1.00       0        0         1         0  TRUE
3    0.4 -1.386219 2.00       0        0         0         1 FALSE
4    0.4 -1.386219 1.75       1        0         0         1  TRUE
5    0.2  7.824221 4.00       0        1         0         0  TRUE
\end{Soutput}
\end{Schunk}

\subsection{Designing and applying a treatment plan with no target}

If there is no target to be predicted (or no outcome variable) \pkg{vtreat} can still be used to prepare data.  In this case the preparation is limited to column cleaning, indication of missing values and production of dummy/indicator variables.  No-target is not a primary intended use of \pkg{vtreat} but is supplied as a convenience for users who may have this data preparation need.

The procedure is as follows:

\begin{Schunk}
\begin{Sinput}
R> treatments <- designTreatmentsZ(d, c('x', 'z'))
\end{Sinput}
\end{Schunk}
\begin{Schunk}
\begin{Sinput}
R> print(treatments$scoreFrame[, scols])
\end{Sinput}
\begin{Soutput}
    varName sig extraModelDegrees origName  code
1    x_catP   1                 2        x  catP
2         z   1                 0        z clean
3   z_isBAD   1                 0        z isBAD
4  x_lev_NA   1                 0        x   lev
5 x_lev_x_a   1                 0        x   lev
6 x_lev_x_b   1                 0        x   lev
\end{Soutput}
\end{Schunk}

The variable types produced when there is no predictive target are as follows.

\begin{itemize}
\item \code{clean} : a numeric variable passed through with all NA/NaN/infinite values replaced with either zero or mean value of the non-NA/NaN/infinite examples of the variable.
\item \code{is\_Bad} : a companion to the \code{clean} treatment.  \code{is\_Bad} is an indicator that indicates a value replacement has occurred.
\item \code{lev} : a 0/1 indicator indicating a particular value of a categorical variable was present.  For example \code{x_lev_x.a} is 1 when the original \code{x} variable had a value of ``a''.  These indicators are essentially variables representing explicit encoding of levels as dummy variables. In some cases a special level code is used to represent pooled rare values.
\item \code{cat\_P} : the prevalence (frequency) of each categorical level in the training data.  This indicates if the original level was rare or common.
\end{itemize}

And we can prepare \code{data.frames} as follows.

\begin{Schunk}
\begin{Sinput}
R> dTreated <- prepare(treatments, d, pruneSig=NULL)
R> print(dTreated)
\end{Sinput}
\begin{Soutput}
  x_catP    z z_isBAD x_lev_NA x_lev_x_a x_lev_x_b
1    0.4 0.00       0        0         1         0
2    0.4 1.00       0        0         1         0
3    0.4 2.00       0        0         0         1
4    0.4 1.75       1        0         0         1
5    0.2 4.00       0        1         0         0
\end{Soutput}
\end{Schunk}

\subsection{Cross-frames and nested models}
\label{sec:crossframes}

In all cases above we have produced a treatment plan called \code{treatments} and a
prepared version of our original \code{data.frame} called \code{dTreated}.  Each of these
examples has the undesirable property that the exact same data \code{d} was used to collect
statistics to design the data preparation plan and then used when applying the transformation.
This can lead to its own form of undetected over-fitting (see Section~\ref{sec:nestedmodelbiastheory} for discussion)
and is an issue worth avoiding.

One way to avoid the issue is to reserve a fraction of data for only
the treatment design phase and to not re-use that data for any other
modeling or evaluation step. This is demonstrated in
Section~\ref{sec:useofcalset}.  This procedure can be statistically
inefficient, so it is important to have an alternative which we call
``cross frames'' or ``simulated out of sample frames''.

\subsubsection{Designing and applying a treatment plan and a simulated out of sample frame for a numeric target}

Instead of the \code{designTreatmentsN}/\code{prepare} sequence we produce our treatments and prepared
\code{data.frame} in one step as shown below.

\begin{Schunk}
\begin{Sinput}
R> cfe <- mkCrossFrameNExperiment(d, c('x', 'z'), 'yN')
R> treatments <- cfe$treatments
R> dTreated <- cfe$crossFrame
\end{Sinput}
\end{Schunk}

At this point we have a \code{treatments} object (with a \code{scoreFrame} value inside it) and a prepared or treated copy
of our original data frame \code{d}.  We can use the prepared data, \code{dTreated}, to fit a model to predict the outcome \code{yN}.
We can use the treatment plan \code{treatments} to prepare future application data as
before. The merit of using \code{mkCrossFrameNExperiment} is that each row of the treated training data frame is produced by
cross-validation: the treatment plan that produces each row was designed \textit{excluding} that row. See Section~\ref{sec:nestedmodeling}
for a discussion. This simulates what will be true for future application data (which is also not involved in the formation of the treatment plan) and
decreases the issue of nested modeling bias.

By default \code{mkCrossFrameNExperiment} uses 3-fold cross-validation; this is controlled by the parameter \code{ncross}.

A few things to note:

\begin{itemize}
\item The \code{cfe$treatments} treatment plan is estimated using all of the data \code{d}.
\item \code{cfe$crossFame} doesn't necessarily equal \code{prepare(cfe$treatments, d, ...)}.
\item Due to the possibility of re-sampling \code{dTreated} may not be a deterministic function of
  \code{d}.  The re-sampling method used is given as the string \code{cfe$method} and the exact re-sampling plan is returned as \code{cfe$evalSets}.
\end{itemize}

\subsubsection{Designing and applying a treatment plan and a simulated out of sample frame for a categorical target}

For a categorical target the cross-frame procedure demonstrated below:

\begin{Schunk}
\begin{Sinput}
R> cfe <- mkCrossFrameCExperiment(d, c('x', 'z'), 'y', TRUE)
R> treatments <- cfe$treatments
R> dTreated <- cfe$crossFrame
\end{Sinput}
\end{Schunk}

\subsection{Sampling controls}
\label{sec:samplecontrols}

The \code{mkCrossFrame*Experiment} methods are centered around building a re-sample of the original
data frame.  The \code{designTreatments*} methods also use this technique to score
complicated variables (those with more than zero hidden degrees of freedom in the \code{scoreFrame})
and get more reliable significance estimates.
The user may need some control of the structure of this re-sampling.
To allow control \code{mkCrossFrame*Experiment} methods take an argument called
\code{splitFunction} which accepts a user supplied function with signature
\code{function(nRows, nSplits, dframe, y)}, where:

\begin{itemize}
\item \code{nRows} is the number of rows you are trying to split.
\item \code{nSplits} is the number of split groups you want. This argument is ignored when doing one-way holdout (leave-one-out cross validation).
\item \code{dframe} is the original data frame, which can be used to identify groups and other features that influence the re-sampling.
\item \code{y} is the outcome, given as numeric values; \pkg{vtreat} converts categorical targets to an indicator before calling the re-sampling function. \code{y} can be used for stratification.
\end{itemize}

The function should return a list of lists.  The $i$th element should have slots \code{train} and \code{app}, where
\code{[[i]]$train} designates the training data used to fit the model that evaluates the data designated by \code{[[i]]$app}.

The structure returned by a \code{splitFunction} is easiest to show through an example.
Here we split a hypothetical 3-row data frame into 3 partitions, using the \code{vtreat::oneWayHoldout} function. The remaining arguments to
the function are \code{NULL} because \code{vtreat::oneWayHoldout} ignores them.

\begin{Schunk}
\begin{Sinput}
R> str(vtreat::oneWayHoldout(3, NULL, NULL, NULL))
\end{Sinput}
\begin{Soutput}
List of 3
 $ :List of 2
  ..$ train: int [1:2] 2 3
  ..$ app  : int 1
 $ :List of 2
  ..$ train: int [1:2] 1 3
  ..$ app  : int 2
 $ :List of 2
  ..$ train: int [1:2] 1 2
  ..$ app  : int 3
 - attr(*, "splitmethod")= chr "oneway"
\end{Soutput}
\end{Schunk}

Each partition designates the indices of a two-row ``training set'' that can be used to build a model to score the data in the one-row ``application set''. The training and application sets are complementary to each other. The above is a leave-one-out cross validation plan. Internally, \pkg{vtreat} uses this structure to specify the construction of the cross frames in \code{mkCrossFrame*Experiment}.

\code{vtreat} supplies a number of cross validation split/plan implementations:

\begin{itemize}
\item \code{kWayStratifiedY}: k-way y-stratified cross-validation. This is the \code{vtreat} default splitting plan.
\item \code{kWayCrossValidation}: k-way unstratified cross-validation
\item \code{makekWayCrossValidationGroupedByColumn}: k-way y-stratified cross-validation that preserves grouping (for example, all rows corresponding to a single customer or patient, etc). This is a generator that returns the complex splitting plan function, and only recommended when absolutely needed.
\item \code{oneWayHoldout}: jackknife, or leave-one-out cross-validation.\footnote{Note leave-one-out cross-validation can leak the expected value of y, so should not be a preferred method in nested modeling situations.}
\end{itemize}

\subsection{Variable significance}
\label{sec:sigop}

The \pkg{vtreat} treatment design methods report variable significances in the
treatment object's \code{scoreFrame}.  Significances are also used as
pruning thresholds in various places, including \code{prepare}'s required
argument \code{pruneSig} that (if not \code{NULL}) is used to prune
variables.

There are issues when using the same
data to produce a treatment and to score the quality of the treatment
(please see Section~\ref{sec:nestedmodelbiastheory}).
For ``simple variables'' (those which have \code{extraModelDegrees} $\le 0$), 
variable significances are computed naively: directly on the data used to design them.
For complex variables (those which have \code{extraModelDegrees} $> 0$), 
\pkg{vtreat} uses a simulated out of sample cross-validation procedure to estimate
the variable significance (see Section~\ref{sec:nestedmodelbiastheory}).

\subsubsection{Significance pruning}

Most current machine learning methods can be overwhelmed by large numbers of
irrelevant variables, 
even those methods that include cross-validation, regularization, and early
stopping as part of their design.\footnote{
Please see \cite{badbayes}
for worked examples of non-pruned useless variables overwhelming naive Bayes, decision trees, 
logistic regression and random forests.} Therefore, variable pruning before modeling is
often advisable.

Because there are several candidate variables to evaluate, variable evaluation and significance-based
pruning suffers from the  multiple comparisons problem. A pragmatic solution is to
set the pruning threshold as the most permissive (largest) value that
compensates for this issue: one over the number of candidate variables.
Please see Section~\ref{sec:varchoiceheuristic}
for details.

\subsection{Rare level options}

The methods \code{designTreatments*} and \code{mkCrossFrame*Experiment} provide some extra parameters
to control rare level processing for categorical variables.

\begin{itemize}
\item \code{minFraction}:  only levels of a categorical variable that occur with at least a \code{minFraction}
frequency in the treatment design data are eligible to be re-encoded as new indicator or dummy variables.  This
option defaults to $0.02$ so that each categorical will by default only expand into a limited number of new
indicator variables.
\item \code{rareSig}: levels that achieve a statistical significance \textit{score} higher (less significant) than \code{rareSig} are suppressed and not eligible to contribute modeling effects.  Per-level significance tests can be expensive, and \code{rareSig} defaults to \code{NULL} meaning ``off.''
\item \code{rareCount}: Levels that occur no more than \code{rareCount} times during training are eligible to be re-coded as a common ``rare level'' symbol unless they fail a statistical test driven by \code{rareSig}.  This feature is described in a vignette, but is mostly inferior to using a \code{catP} variable (or using an interaction of a \code{catN} or \code{catB} with a \code{catP}).  The default is $0$, meaning this control defaults to ``off.''
\item \code{smFactor}: Number of pseudo-observations to add as a Laplace smoothing factor (to reduce the range of predictions of rare levels).
\end{itemize}

Currently \pkg{vtreat} does not supply per-variable settings of these controls; one setting is used for the entire
process.

\subsection{Parallelism}

For large datasets \pkg{vtreat} treatment design can take some time (though usually much less time than the modeling steps that follow).
To help mitigate this, \pkg{vtreat} operations take an argument \code{parallelCluster}.  This argument can be a parallel operation cluster
built by the package \code{parallel} or \code{snow}.  If given such an argument \pkg{vtreat} will schedule operations using
\code{parallel::parLapplyLB}.  \pkg{vtreat}'s use of parallelism is compatible with socket clusters, so it should work on all \proglang{R}
architectures.

A typical use of parallelism is given below.

\begin{Schunk}
\begin{Sinput}
R> ncore <- 2
R> parallelCluster <- parallel::makeCluster(ncore)
R> cfe <- mkCrossFrameNExperiment(d, c('x', 'z'), 'yN', 
+     parallelCluster=parallelCluster)
R> parallel::stopCluster(parallelCluster)
\end{Sinput}
\end{Schunk}

\subsection{Scaling}

\code{prepare} and \code{mkCrossFrame*Experiment} both accept an
argument called \code{scale} which defaults to \code{FALSE}.  If set
to \code{TRUE}, all derived variables or columns are rescaled to
``\code{y}-units.'' See \cite{yawarescaling} for details.

The scaling feature is particularly useful as a pre-processing step
for principal components analysis, clustering, and general dimension reduction\footnote{
In such applications it makes sense to center and scale the dependent or
target variable $y$ to be mean zero and variance one before treatment design.}.

\section[Current limitations of vtreat]{Current limitations of \pkg{vtreat}}
\label{sec:limitations}

In this section we briefly identify some of the limitations of the current \pkg{vtreat} implementation.
\pkg{vtreat} has recently added big data capabilities (via \code{vtreat::rquery\_prepare()}) and multi-class
classification extensions (through \code{vtreat::mkCrossFrameMExperiment()}). However there are still 
directions for potential improvement:

\begin{itemize}
\item \pkg{vtreat} does not currently help look for interactions involving high cardinality categorical variables, beyond converting common levels to indicators.
\item \pkg{vtreat}'s missing value treatment is isolated and point-wise, not conditioned other variables or distributional.
\end{itemize}

\section{Final remarks}
\label{sec:finalRemarks}

Data preparation using \pkg{vtreat} can improve the performance of predictive models in production. This is a strong argument to add \pkg{vtreat} to the predictive analytics work-flow.

\bibliography{vtreat}

\end{document}